# A Model of the Commit Size Distribution of Open Source

Carsten Kolassa[1], Dirk Riehle[2] and Michel A. Salim[3]

[1] RWTH Aachen `Carsten@Kolassa.de`
[2] Friedrich-Alexander-University Erlangen-Nürnberg `dirk@riehle.org`
[3] Friedrich-Alexander-University Erlangen-Nürnberg `michel@sylvestre.me`

**Abstract.** A fundamental unit of work in programming is the code contribution ("commit") that a developer makes to the code base of the project in work. We use statistical methods to derive a model of the probabilistic distribution of commit sizes in open source projects and we show that the model is applicable to different project sizes. We use both graphical as well as statistical methods to validate the goodness of fit of our model. By measuring and modeling a fundamental dimension of programming we help improve software development tools and our understanding of software development.

## 1 Introduction

Free/libre/open source software (FLOSS) has been adopted widely by industry in recent years. In 2008, 85% of all enterprises were using open source software [8]. A 2010 study estimated that 98% of all enterprises were using open source software [24].

Given the significance of open source, it is surprising that there are few representative statistical analyses of open source projects, and that there are no high quality models of the fundamental dimensions of programming in open source projects.

In this paper we present a model of one important dimension of programming in open source software development, the distribution of the sizes of code contributions made to open source projects. This so-called commit size distribution describes the probability that a given commit is of a particular size.

A commit is an individual code contribution of a developer. [14] show that lines of code are a good proxy for work spent on that code. Hence, a commit is a basic unit of work performed by a developer.

The commit size distribution is therefore a model of fundamental units of work performed in open source programming.

Understanding the work performed in open source programming is helpful for building better software development tools and understanding software development in general. Moreover, case studies suggest that open source is similar to closed source in terms of growth, project complexity or modularity [19]. Thus the results of this paper are likely to apply not only to open source but to closed source as well.

The contributions of this paper are:

1. A high quality analytical model of the commit size distribution of open source.
2. An in-depth validation of the model using appropriate statistical measures.
3. A comparison of commit size distributions of different project sizes.

The rest of the paper is organized as follows. Section 2 describes the necessary terms. Section 3 defines and analyzes the commit size distribution. Section 4 discusses the potential threats to validity and section 5 discusses prior and related work. We consider potential extensions in section 6, and present our conclusions in section 7.

## 2 Commit Sizes

A software project is typically developed in multiple iterations, in a series of changes to its artifacts, for instance, code, documentation, or artwork. If a project is managed using a *version control system* (also known as a *source code management* system), these changes are organized into sets known as *commits*.

In this paper we address programming, hence we are only concerned with source code commits. We measure the commit size in terms of *lines of code* (LoC). We distinguish between source code lines, comment lines, and empty lines. We use the following definitions:

1. a source code line (SLoC) is one line of program code,
2. a comment line (CL) is a line consisting only of comments,
3. an empty line contains only whitespace, and
4. a line of code (LoC) is either a source code line or a comment line.

Measuring the size of a commit is a non-trivial task. The main tool for assessing commit sizes is the "diff" tool which tells the user which lines have been added and which lines have been removed. Unfortunately, a diff tool cannot iden-

$$lower\_bound(a, r) = max(a, r)$$
$$\text{full overlap, highest number of changed lines} \quad (1)$$
$$upper\_bound(a, r) = a + r$$
$$\text{no overlap, no changed lines in diff chunk} \quad (2)$$
$$diff\_chunk\_size(a, r) = \frac{(lower\_bound(a, r) + upper\_bound(a, r))}{2}$$
$$\text{mean value of lower and upper bound} \quad (3)$$

Fig. 1: Equations used to compute a commit's size from input lines added and removed

tify with certainty whether a line was changed, because a changed line is always counted as one line removed and one line added. However, a changed line should count as one line of work, while an added and a separately removed line of code should count as two lines of work.

[4] developed an algorithm for identifying changed lines of code from added and removed lines of code. They use the Levenshtein distance algorithm which is a metric for measuring the distance between two strings. While helpful this approach has one major disadvantage: it is computationally expensive and does not scale to large amounts of source code. Because our analysis covers about 30% of all of open source code at its time, we need another approach.

We use the sample data from Canfora et al. to derive a simple function for estimating diff chunk sizes where the diff chunk size is a function of two variables: lines of code added and lines of code removed. This function provides a statistically valid estimate for the size of a given diff output as shown in [13]. However, our evaluation based on Canfora's sample data revealed that the regression performs only trivially better than estimating the diff chunk size by taking the mean of the minimum possible and the maximum possible sizes. Thus, a more plausible and unbiased algorithm for estimating the statistically expected value is simply to take the mean of the minimum and maximum possible values.

Figure 1 provides the necessary equations. In these equations, $a$ represents the number of lines added and $r$ represents the number of lines removed according to the diff tool.

In this paper we compute commit sizes by adding up the diff chunk sizes computed using equation 3 of figure 1. A diff chunk size is the size of the diff of one file in the commit. After calculating the size of every commit in our data set we compute the commit size distribution. The commit size distribution describes the relative likelihood that a commit has a particular size. The commit size distribution of some commit population is the distribution of the probabilities (or number of occurrences) of all possible commit sizes.

## 3 Commit Size Distribution

### 3.1 Data Source and Research Method

This paper uses the database of the Ohloh.net open source project index. Our database snapshot is dated March 2008. It contains 11,143 open source projects with a total of 8,705,118 commits. [6] estimates that there were 18,000 active open source projects in September 2007 worldwide. The total number of projects is much larger, but most open source projects are not active and by our activity definition have to be excluded. We use the same definition of "active project" as Daffara: A project is active at a given point in time if the number of commits in the preceding 12 months is at least 60% of the number of commits in the 12 months before that. Using this definition our data set contains 5,117 active open source projects. We therefore estimate that our database contains about 30% of all open source projects considered active in March 2008.

Our analysis is descriptive: we are discovering existing characteristics in our data rather than starting off with a hypothesis and attempting to invalidate or validate it. We provide details not only of our final findings but also of the attempted distributions that did not fit. We also split our analysis along project sizes and provide the characteristics of commit size distributions by project size.

### 3.2 Measurements

We determined the total commit size distribution of our open source sample population using the definition of section 2. In statistics a distribution can be represented as a *probability distribution function* (PDF) or a *cumulative distribution function* (CDF). The PDF in our case describes the relative likelihood that a commit of a certain size occur at a given point. The CDF can be computed by integrating the PDF. Integrating the PDF over an interval provides the probability that a commit is of the size determined by the interval boundaries. For example, integrating over the interval $[1, 10]$ provides the probability that a commit has between 1 and 10 lines of code, 1 and 10 included.

The empirical result of our measurements is the *empirical probability distribution function* (EPDF) as shown in figure 2. The EPDF is a density estimation based on the observed data. It describes the probability that a certain commit has a certain commit size. The EPDF is not a closed model, it is just a representation of the observed data. The statistical key characteristics are shown in table 1.

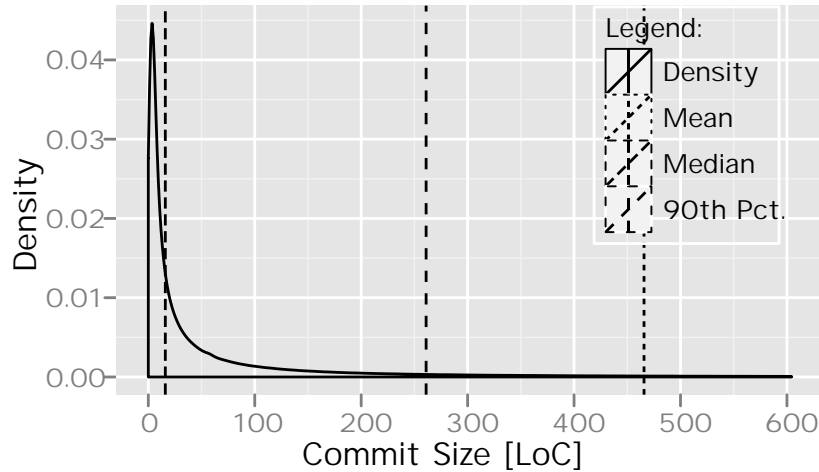

Fig. 2: The EPDF of the commit size distribution up to the 95th percentile of about 30% of all active open source projects (March 2008)

| Key Parameter | Value |
| --- | --- |
| Mean | 465.72 |
| Median | 16 |
| 90th percentile | 261 |
| 95th percentile | 604.5 |

Table 1: Statistical key characteristics of the open source commit size distribution

The form of the EPDF, in particular that it is monotonically falling suggests the following possible probability distributions:

- Biexponential
- Exponential
- Generalized Pareto
- Pareto
- Simple Power Law
- Weibull
- Zipf's

To determine the analytically closed form of the PDF and CDF we calculated the *empirical cumulative distribution function* (ECDF) and fitted the different possible distributions to the ECDF based on Newman's advice to choose the CDF for analytical purposes [17]. This approach is robust and allows us to use different regression techniques without binning. Thus we prevent the introduction of biases and information loss that comes with binning. We then fitted the different distributions (see enumeration above).

After reviewing the different fits and the residual plots as well as the P-P plots ("P" stands for percentile) we found that the Generalized Pareto Distribution (GPD) provides the best fit.

The GPD is broadly applicable and incorporates both exponential and Pareto distributions when certain parameters are fixed [15].

The Generalized Pareto Distribution is difficult to fit using the maximum likelihood approach, as the location parameter is unbounded (see [22] and [21]). We therefore decided to use a least square fit on the ECDF. The location parameter is chosen manually, attempting to fit the other two parameters with increasing values of location in increments of 0.5 (the granularity of commit size estimates, since they are averages of two integral values). We find that a value of 0.5 minimizes the difference between CDF and ECDF at the mode commit size of 1.

$$f(x) = \begin{cases} \frac{1}{\sigma}(1+\xi\frac{x-\theta}{\sigma})^{-1-\frac{1}{\xi}} & \text{for } \xi \neq 0 \\ \frac{1}{\sigma}exp(-\frac{(x-\theta)}{\sigma}) & \text{for } \xi = 0 \end{cases} \quad (4)$$

Fig. 3: PDF formula for the Generalized Pareto Distribution.

$$F(x) = \begin{cases} 1 - (1 + \frac{\xi(x-\theta)}{\sigma})^{-1/\xi} \text{ for } \xi \neq 0 \\ 1 - exp(-\frac{x-\theta}{\sigma}) \text{ for } \xi = 0 \end{cases} \quad (5)$$

Fig. 4: CDF formula for the Generalized Pareto Distribution.

The result of our fit is the CDF of the commit size distribution in closed form, which is shown next to the ECDF in figure 5.

The parameters of the Generalized Pareto Distribution are shown in table 2, the equations for the Generalized Pareto Distribution are shown in figure 3 and 4. Where $\theta$ is the location parameter, it controls how much the distribution is shifted. $\sigma$ is the scale parameter it controls the dispersion of the distribution, while $\xi$ is the shape parameter which controls the shape of the generalized pareto distribution [5].

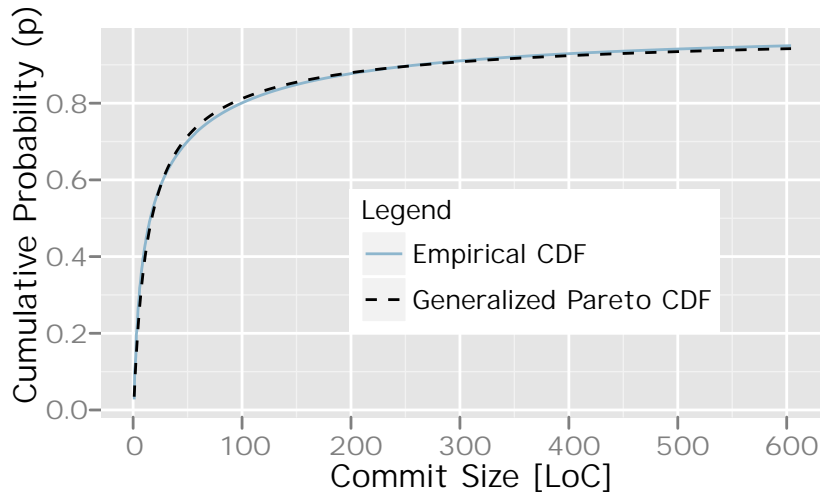

Fig. 5: The Generalized Pareto Distribution of the CDF up to the 95th percentile of about 30% of all active open source projects (March 2008)

| Parameter | Value |
| --- | --- |
| $\xi$ (xi) / Shape | 1.4617 |
| $\theta$ (theta) / Location | 0.5 |
| $\sigma$ (sigma) / Scale | 13.854 |

Table 2: Model parameters of Generalized Pareto Distribution as calculated from least squares

Figure 6 shows the P-P plot that compares our model to the empirical data. A P-P plot is a graphical method to compare two probability distributions by plotting their percentiles against each other; here, we compare the percentiles of our model (i.e. the CDF) to the percentiles of the empirical data (i.e. the ECDF).

As can be observed by examining the CDFs of our model and empirical data, both distributions are long-tailed; thus, per [10], the P-P plot is more appropriate than the more familiar Q-Q (quantile-quantile) plot.

To show the goodness of our fit we also compute quantitative measures such as R-square and Pearson's R as shown in table 3.

| Parameter | Value |
| --- | --- |
| R-square on CDF | 0.9949 |
| Pearson's R on CDF | 0.99755 |

Table 3: Goodness of Fit Indicators calculated up to the 95th percentile.

### 3.3 Comparison by Project Size

We also want to understand how the commit size distribution varies by project size. One might hypothesize that small projects are different from medium sized and large projects. However, we found that the GPD not only fits when analyzing all projects, it also fits to subsets of different sizes in terms of number of developers.

We classify the projects into small, medium, and large sized projects based on the number of involved developers. [3] provide an analysis of the number of developers in a random sample of projects included in the Debian GNU/Linux distribution. We use their proposed partitioning to group our projects accordingly (see table 4).

We can now measure how the commit size distribution correlates with project sizes. We found that the commit size distribution of small, medium, and large projects are also best characterized as generalized Pareto distributions. Figures 10, 11, and 12 show the cumulative distribution functions respectively.

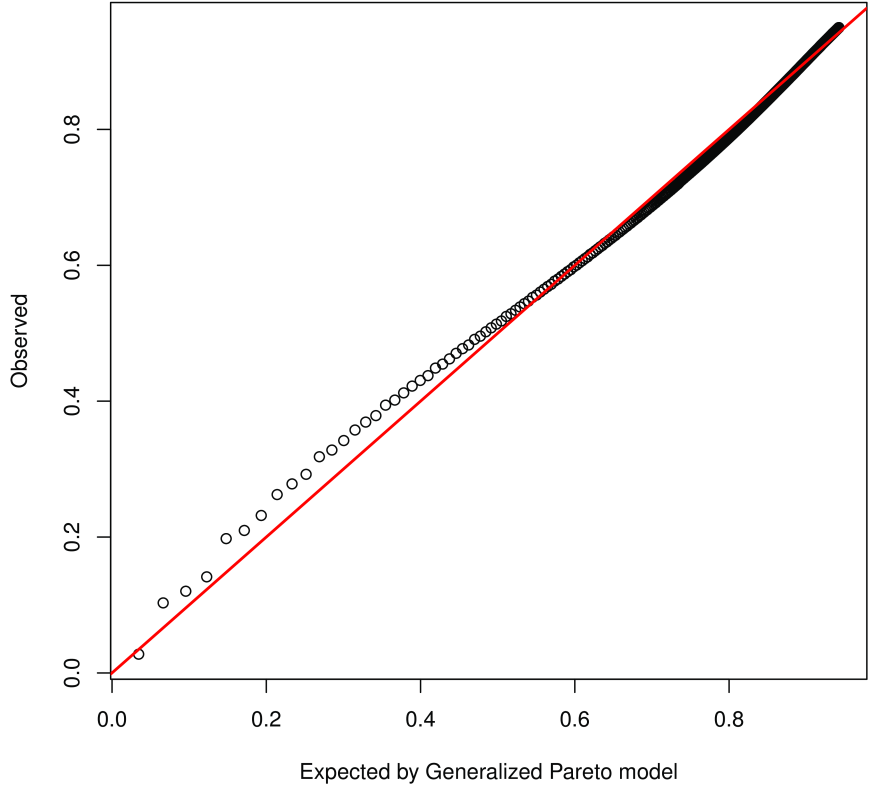

Fig. 6: P-P plot comparing the percentiles of the model and the empirical percentiles

| Parameter | Minimum number of developers | Maximum number of developers |
|---|---|---|
| Small | 1 | 5 |
| Medium | 6 | 47 |
| Large | 48 | $\infty$ |

Table 4: Project size boundaries

The parameters of these distributions (see Table 5) are close to the parameters of the total distribution. Figure 13 compares the EPDFs of the different subsets. After comparing the plots and the model parameters we came to the conclusion that the location parameters is invariant to the number of developers in a project. The shape parameters have no obvious correlation with the number of developers and the differences are small, while the scale parameter falls as the number of developers increases. A possible explanation for this is that when more and more developers join a project the average commit size goes down to prevent merge conflicts. Another explanation is the observation of [23] that small patches are more likely to be accepted then large ones; we posit that this affects larger projects more since they are more likely to have a formalized code review process.

The effect is not very big; in fact, in the region of commit sizes with the most pronounced difference (commits smaller than 13 LoC), the difference in proportion of commits in this category between large and small projects is 6.03 %. For commits in this region, both models have errors (the difference between the respective model and the empirical data) smaller than 3%.

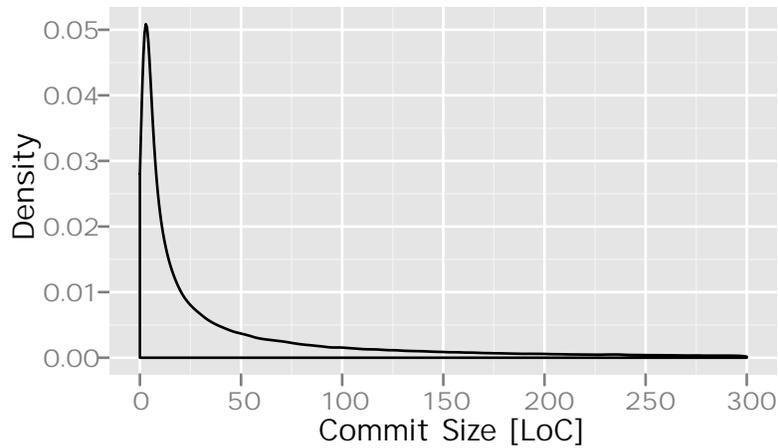

Fig. 7: EPDF for small projects.

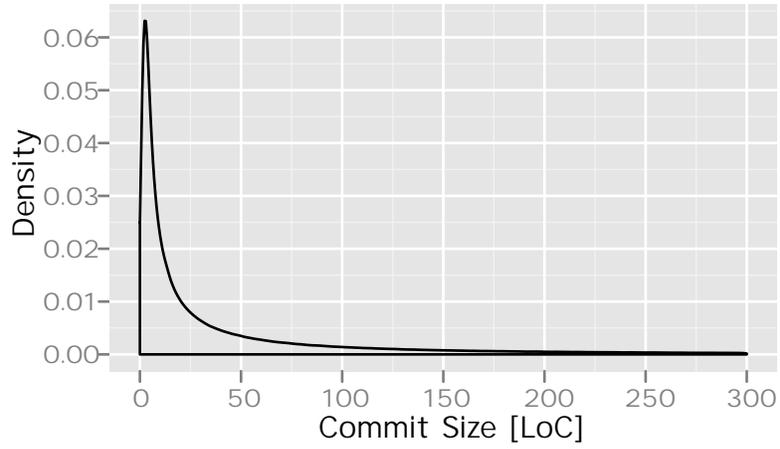

Fig. 8: EPDF for medium projects.

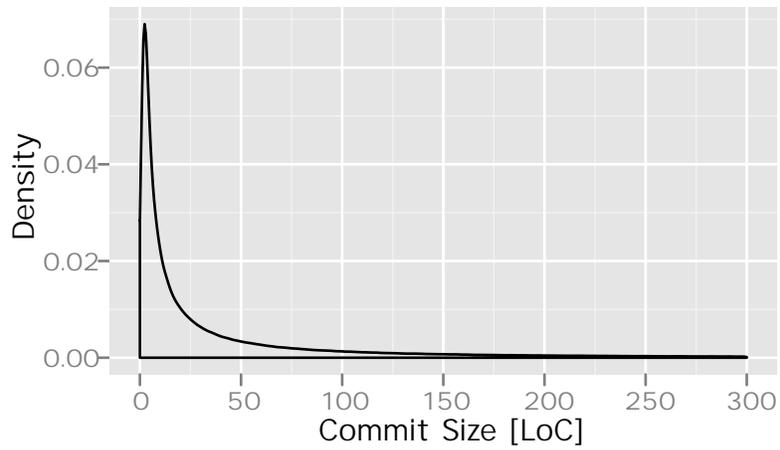

Fig. 9: EPDF for large projects.

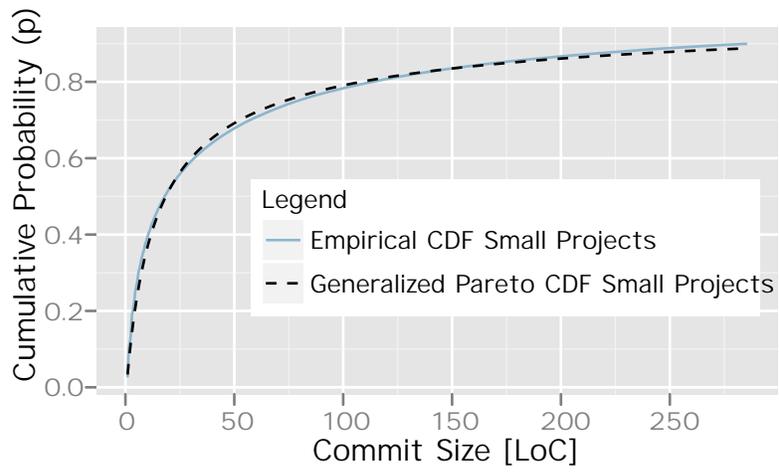

Fig. 10: Generalized Pareto Distribution as CDF of small open source projects (March 2008).

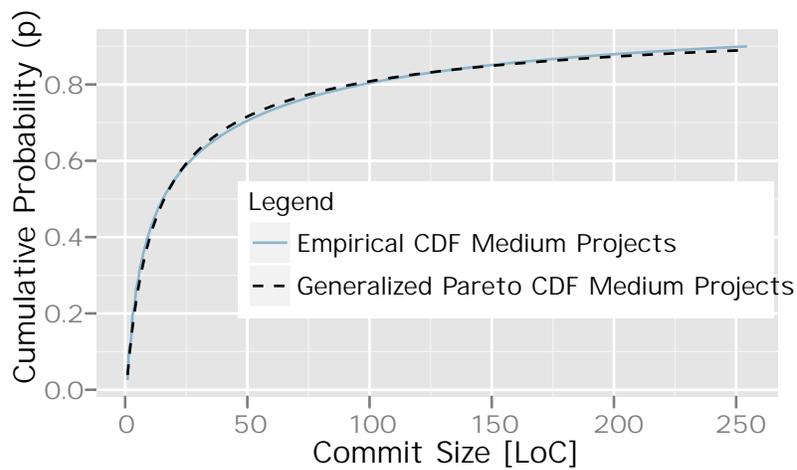

Fig. 11: Generalized Pareto distribution as CDF of medium open source projects (March 2008).

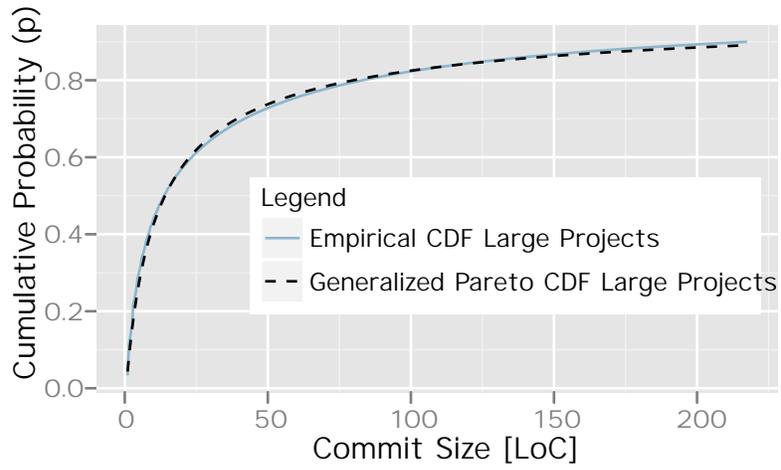

Fig. 12: Generalized Pareto distribution as CDF of large open source projects (March 2008).

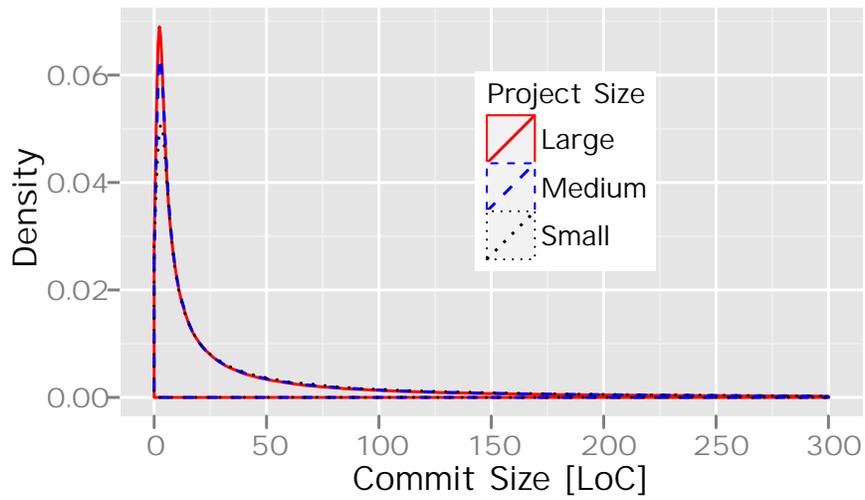

Fig. 13: EPDF for different project sizes for easy comparison.

| Parameter | Small projects | Medium projects | Big projects |
|---|---|---|---|
| $\xi$ (xi) / Shape | 1.5969 | 1.6008 | 1.5708 |
| $\theta$ (theta) / Location | 0.5 | 0.5 | 0.5 |
| $\sigma$ (sigma) / Scale | 14.249 | 12.199 | 10.822 |

Table 5: Model parameters of the Generalized Pareto Distribution as calculated for different project sizes.

# 4 Threats to validity

## 4.1 Poor support for multiple enlistments

The Ohloh 2008 dataset has limited support for multiple enlistments [18]. Projects that have changed their SCM, either moving it to a new URL or transitioning to newer technology (e.g. CVS $\rightarrow$ SVN $\rightarrow$ Git or Mercurial) thus face either having their commits listed multiple times (resulting in older commits being given too much weight). But in practice this is not a problem because there is no difference in the distribution of old and new commits [7].

## 4.2 Model errors

The slight over- and undercutting is an indication for a systematic error in our model. We cannot explain this systematic error, but the goodness of fit calculations and the P-P plot show that it is sufficiently small. That is the reason why we do not address it further. There might be a second underlying distribution that is responsible for this error but we have not been able to determine it. We also think that removing that error would come at the risk of overfitting our model.

## 4.3 Bias of the Ohloh data

As mentioned earlier we rely on the Ohloh data for our analysis. The Ohloh data has two self-selection biases:

Projects that die very early probably never make it into the Ohloh database and are therefore underrepresented in our analysis. Projects from non-English speaking countries are also less likely to be included in Ohloh.

We think the first bias is not an issue because for such projects it would be difficult to derive a statistically significant measure as there are almost no commits yet. We also think that the language bias is unproblematic because we do not think that there are differences in the commit size distribution whether a project is done in an English- or non-English-speaking environment.

# 5 Related work

We previously presented a preliminary analysis of the commit size distribution [2]. Compared to this work our current analysis adds a closed model and a validation of this model as well as an analysis of the commit size distribution by project size.

[1] present an analysis of "a typical commit" using the version history of 9 open source projects. They mostly focus on the number of files changed (and how), but also provide chunk and line-size data. They compute line size changes by adding lines added and removed, thus overestimating sizes by ignoring changed lines of code. Still, they find "quite small" commit sizes without giving more details. Interestingly, they find a strong correlation between diff chunk and size. Alali et. al.'s 9 projects are large well-known open source projects. In contrast to Alali we focus solely on commit size, use a more precise measure and compute a derived function, the commit size distribution, on a more representative sample rather than 9 selected projects.

[20] analyze the impact that small changes have on various quality attributes of the software under consideration. Their data is derived from a single large closed source project. They find that one-line changes represent the majority of changes during maintenance, which is in line with our results. [12] analyze 2,000 large commits from 9 selected open source projects and they find that small commits are more corrective while large commits are more perfective. Unfortunately, the authors do not discuss as to what extent their results might be representative of open source. [23] look at the patch submission and acceptance process of two open source projects. They find that small patches are more likely to get accepted into the code base than large patches. An obvious reason may be, that smaller patches are easier to review than large patches which, if not handled quickly, get harder to review and accept with time. While not representative, Weißgerber's observation is interesting to us, as it might explain why the commit size distribution is skewed towards small commits, and why this skewness is more pronounced in larger projects.

The analysis of code repositories for various purposes is an important research area that has given birth to the annual Mining Software Repositories conference series, usually co-located with ICSE [11]. A 2009 IEEE Software special issue on Mining Software Archives [16] was followed by a symposium of the same name in 2010. Ghezzi and Gall propose not only to undertake such research but to provide a platform that allows for the distributed composition of services for such analysis work [9].

Our research has one key distinguishing feature when compared to other open source data analysis research: The size of our sample population is much larger than any other published data set and brings us close to being representative of open source.

## 6 Future work

### 6.1 Study of proprietary software

We would like to extend our analysis to that of proprietary software projects. In order to do this we require access to commit statistics of proprietary software projects, and this in turn requires collaboration with software vendors to get access to their statistics.

### 6.2 Validation by accessing software repositories

For the open-source projects that we analyze it would be desirable to validate our findings by picking several key projects, mining their revision history ourselves (rather than depending on the Ohloh statistics) and comparing the results to the same statistics computed over the Ohloh data (for the same projects).

### 6.3 Extended analysis

Some analysis are not possible with the Ohloh data, and like those above require direct access to the software repositories:

Certain version control systems, like Git allow a commit to have an author, multiple signatories, and a committer. With certain others (e.g. SVN) it's not built-in, and projects have to resort to informal conventions for marking commit authorship if the author does not have commit access. Thus it is not possible to reliably reconstruct this data.

With direct access we could better characterize projects based on write access to code – whether BSD-style (a core team with commit bit can touch any part of the codebase; centralized development), Linux-style (a hierarchical system with lieutenants in charge of certain parts of the codebase; distributed development, changes can still be made to the entire tree, but commits tend to be accepted only if they are within the developer's competence), commercial open source (most development is done by paid employees; external fixes might be accepted but are committed by a paid employee)

## 7 Conclusions

This paper shows that small commits are much more likely than large commits with 50% being 16 lines of code or less.

The actual commit size distribution of open source is best modeled by a Generalized Pareto Distribution and we have found the same kind of distribution fits for different project sizes, with the likeliness of small commits increasing with the number of developers.

The fact that it is a Pareto Distribution, which is a distribution with a long tail, also shows that large commits happen although they are less likely.

The empirical knowledge gained from actually measuring the commit size distribution is the first step to creating hypotheses for future research to improve software development tools. It can be used as a benchmark to compare projects as well.

The mathematical model presented in this paper is one step towards a more precise model of software development. It is also important for developing new software development methodologies and to develop a general model of software development.